\documentclass{spie}%
\usepackage{amsfonts}
\usepackage{amsmath}
\usepackage{amssymb}
\usepackage{graphicx}%
\usepackage{subfigure}
\input{tcilatex}

\begin{document}

\title{Microjoule mode-locked oscillators: issues of stability and noise}
\author{Vladimir L. Kalashnikov\supit{a} and Alexander A. Apolonski\supit{b}\\\supit{a}Institut f\"{u}r Photonik, TU Wien, Gusshausstr. 27/387, A-1040
Vienna, Austria \\\supit{b}Ludwig-Maximilians-Universit\"{a}t M\"{u}nchen, D-85748 Garching, Germany,  \\ Max Planck Institute of Quantum Optics, Hans-Kopfermann-Str. 1, D-85748 Garching, Germany and
Institute of Automation and Electrometry, SB RAS, Novosibirsk
630090, Russia }
\authorinfo{Further author information:}
\authorinfo{E-mail: kalashnikov@tuwien.ac.at, Telephone: +43-1-58801-38743}
\maketitle

\begin{abstract}
In this work, for the first time to our knowledge, stability and
noise of a thin-disk mode-locked Yb:YAG oscillator operating in both
negative- (NDR) and positive-dispersion (PDR) regimes have been
analyzed systematically within a broad range of oscillator
parameters. It is found, that the scaling of output pulse energy
from 7 $\mu$J up to 55 $\mu$J in the NDR requires a dispersion
scaling from -0.013 ps$^{2}$ up to -0.31 ps$^{2}$ to provide the
pulse stability. Simultaneously, the energy scaling from 6 $\mu$J up
to 90 $\mu$J in the PDR requires a moderate dispersion scaling from
0.0023 ps$^{2}$ up to 0.011 ps$^{2}$. A chirped picosecond pulse in
the PDR has a broader spectrum than that of a chirp-free soliton in
the NDR. As a result, a chirped picosecond pulse can be compressed
down to a few of hundreds of femtoseconds. A unique property of the
PDR has been found to be an extremely reduced timing jitter. The numerical results agree with the analytical theory, when spectral properties of the PDR and the negative feedback induced by spectral filtering are taken into account.

\end{abstract}

\section{Introduction}\label{intro}

High-power ultrafast thin-disk oscillators allow energy-scalable
pico-(ps) and femtosecond (fs) pulse generation at MHz repetition
rates \cite{keller1,keller2}. To date, the over-10 $\mu$J fs-pulses
have been obtained directly from the Yb:YAG thin-disk oscillators
operating in the negative dispersion regime
(NDR)\cite{keller3,neuhaus}. The high-energy pulse generation from a
thin-disk oscillator operating in the positive dispersion regime
(PDR) is achievable, as well\cite{pdr1}. The high-energy fs-pulses
nowadays allow direct experiments on light-matter interactions at
the intensity levels approaching PW/cm$^2$ \cite{haensch,ullrich}.
In particular, high-harmonic generation at such energy levels
promises developing of the table-top VUV/XUV sources, which are of
interest for physics, chemistry, material science, medicine, and
biology. In spite of the chirped-pulse amplification systems (CPA)
\cite{mourou}, the energy-scalable femtosecond oscillators are more
compact, simple, and less expensive. Moreover, the MHz repetition
rates of such oscillators (versus the kHz ones of CPA) reduce
substantially measurement time in the pump-probe experiments as well
as improve signal-to-noise ratio \cite{keller1}.

Energy-scalability issues a challenge of the oscillator stability.
From the soliton area theorem, one may estimate the pulse energy $E$
in the NDR: $ E \propto \sqrt {{{|\beta| P_0 } \mathord{\left/
 {\vphantom {{\beta P_0 } \gamma }} \right.
 \kern-\nulldelimiterspace} \gamma }}
$ [$\beta$ is the group-delay dispersion (GDD) coefficient, $P_0$ is
the pulse peak power, $\gamma$ is the self-phase modulation (SPM)
coefficient]. If one has to confine $P_0$ below some level providing
the pulse stability, the pulse energy in the NDR scales as $ E
\propto \sqrt \beta $. The similar estimation for the law of energy
scaling follows from the energy rate equations for a mode-locked
oscillator \cite{haus}. As a result, the necessary GDD value scales
as square of $E$. In the PDR, the scaling law can be expressed
approximately as $ E \propto \beta ^2 $ \cite{kalash1}. The last
expression suggests, that the scaling properties of the PDR excel
those in the NDR. However, the detailed study of stability needs a
numerical approach. For instance, a departure from the area theorem
appears in the NDR \cite{numer} and the mechanism of pulse
destabilization in the PDR can differ from the CW-excitation
considered in Refs.\cite{kalash1,kalash2}.

An additional issue appears from the requirement of the pulse period
stability imposed by such applications of the oscillators under
consideration as pump-probe experiments, parametric mixing of the
pulses from different oscillators, coherent pulse enhancement in a
resonant cavity, etc. The well-known analytical theory of noises of
mode-locked oscillators \cite{haus2} is based on the soliton
perturbation theory and, thereby, does not deal with the high-energy
regimes, where high-order dissipative nonlinearities contribute, and
with the PDR, where there exists no chirp-free pulse. Moreover,
large gain coefficient and narrow gain bandwidth distinguish the
thin-disk oscillators. As a result, the gain fluctuations affect the
timing jitter deeply \cite{paschotta2}. These effects, which are
beyond the scope of the solitonic model, need a numerical
consideration \cite{paschotta2,paschotta1,paschotta3}.

In this work, the stability thresholds and fluctuations of pulse
group-delay (timing jitter) in a high-energy Yb:YAG thin-disk
oscillator operating in both NDR and PDR are analyzed systematically
on basis of numerical simulations. Two noise sources are considered:
gain fluctuations and quantum noise due to spontaneous emission in
an active medium. It is found, that the scaling properties of the
PDR exceed those of the NDR in the sense, that i) the GDD value
providing the pulse stability is substantially reduced in the PDR
and ii) scales as $\propto \sqrt{E}$ vs. $\propto E$ in the NDR. The
pulse duration is smaller in the NDR, but the spectrum is broader in
the PDR that makes the chirped pulse to be compressible down to
sub-picosecond pulse duration. The group delay caused by the gain
dispersion is sensitive to the gain fluctuations that results in the
timing jitter. However, the timing jitter is substantially reduced
in the PDR as compared to that in the NDR. But in contrast to the
latter, the timing jitter increases with the energy $E$ for the
chirped pulse.

\section{Model of mode-locked oscillator with noise}\label{model}

\begin{figure}
\begin{center}
    \includegraphics[width=6cm]{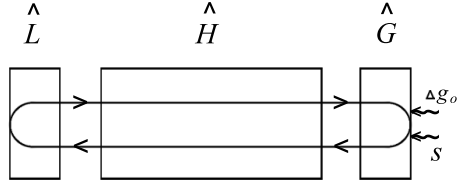}\\
  \caption{Block diagram for the numerical analysis of a Yb:YAG thin-disk oscillator. $\hat L$ is the absorption (both saturable and unsaturable), $\hat H$ is the compensating GDD and the SPM in air, $\hat G$ is the frequency-dependent saturable gain with fluctuating gain coefficient, quantum noise, GDD, and SPM.}\label{fig1}
  \end{center}
\end{figure}

Evolution of time-($t$) dependent slowly-varying field envelope
$a(t)$ inside an oscillator is modeled on basis of the undistributed
map shown in Fig. \ref{fig1}. Here the nonlinear operator

\begin{equation}\label{loss}
    \hat L = \exp \left[ { - \ell  - \frac{\kappa }{{1 + \zeta \left| {a\left( t \right)} \right|^2 }}} \right]
\end{equation}

\noindent describes the net-loss action. The unsaturable loss
coefficient $\ell=$0.07 includes the output loss (=12\%). The
nonlinear (saturable) part of the loss operator describes an action
of a semiconductor saturable absorber with the modulation depth
coefficient $\kappa=$0.005 and the inverse saturation power $\zeta =
{{T_r^a } \mathord{\left/
 {\vphantom {{T_r^a } {E_s^a S_a }}} \right.
 \kern-\nulldelimiterspace} {E_s^a S_a }}
\approx$0.71 MW$^{-1}$, where the absorber relaxation time $T_r^a$
equals to 0.5 ps, the absorption saturation fluency $E_s^a$ equals
to 90 $\mu$J/cm$^2$, and the mode area $S_a$ corresponds to the 1 mm
mode size on an absorber.

The Hamiltonian operator

\begin{equation}\label{hamiltonian}
    \hat H = i\beta \frac{{\partial ^2 }}{{\partial t^2 }} - i\gamma_{air} \left| {a\left( t \right)} \right|^2
\end{equation}

\noindent describes the distributed action of the air nonlinearity
and the GDD of the cavity. The GDD coefficient $\beta$ is variable
($\beta<$0 corresponds to the NDR, $\beta>$0 corresponds to the
PDR). The SPM in air is defined by the coefficient $\gamma_{air}=$3
GW$^{-1}$, that corresponds to the effective mode size of 2.2 mm and
the 11 MHz oscillator repetition rate.

The integro-differential stochastic operator $ \hat G $ means

\begin{equation}\label{gain}
 \hat G\left[ a \right] = i\beta _g \frac{{\partial ^2 }}{{\partial t^2 }}a\left( t \right) - i\gamma _g \left| {a\left( t \right)} \right|^2 a\left( t \right) + \left( {\frac{{\left( {g_0  + \Delta g_0 } \right)\Omega _g }}{{1 + 2\frac{{\int\limits_{ - \infty }^\infty  {\left| {a\left( t \right)} \right|^2 dt} }}{{E_s^g }}}}\int\limits_{ t }^{\infty} {\exp \left[ { - \Omega _g \left( {t' - t} \right)} \right]} a\left( t' \right)dt'} \right) + s\left( t \right)
\end{equation}

\noindent and describes a 200 $\mu$m Yb:YAG thin-disk with the GDD
coefficient $\beta_g=$260 fs$^2$ and the SPM coefficient
$\gamma_g=$0.12 GW$^{-1}$ for the mode size of 2.4 mm. The saturable
gain with the gain coefficient $g_0$ for a small signal and the
saturation energy $ E_s^g  = \frac{{h\nu }}{{\sigma T_r^g }}T_{cav}
S_g  \approx $0.24 mJ ($T_r^g=$1 ms is the gain relaxation time,
$\sigma=$2$\times$10$^{-20}$ cm$^2$ is the gain cross-section,
$T_{cav}$ and $S_g$ are the cavity period and the mode area,
respectively) has the causal Lorentz spectral profile
\cite{oug,belanger,ch} with the width $\Omega_g=$5.3 THz (the gain
bandwidth of 6 nm).

The stochastic (white-noise) value $\Delta g_0$ describes the
initial gain fluctuation (in the limits of $\pm$0.025$g_0$) so that
the value $g_0+\Delta g_0$ differs for each independent simulation
with 10000 cavity round-trips and some fixed set of the parameters
($g_0$, $\beta$, $E_s^g$, etc.). The complex stochastic value $s(t)$
such that \cite{paschotta2}

\begin{equation}\label{noise}
    \left\langle {s\left( t \right)s^* \left( {t'} \right)} \right\rangle  = 2\left(\ell + \kappa \right) \theta\frac{{h\nu }}{\delta t} \delta \left( {t - t'} \right)
\end{equation}

\noindent describes the quantum noise of an active medium ($\theta$
is the enhancement factor due to an incomplete inversion of active
medium, $\delta t$ is the time step in subdividing of time window representing $a(t)$).

The propagation of the field with the complex envelope $a(t)$
through the system shown in Fig. \ref{fig1} is described by

\begin{equation}\label{master}
    a_{k + 1} \left( t \right) = \frac{{\hat L}}{2}\frac{{\hat H}}{2}\hat G\frac{{\hat H}}{2}\frac{{\hat L}}{2}a_k \left( t \right),
\end{equation}

\noindent where $k$ is the cavity round-trip number. The iterative
Eq. (\ref{master}) is solved on basis of the symmetrized split-step
Fourier method on the mesh with the minimum time step $\delta t=$2.5 fs and
the simulation window of $\approx$655 ps. The steady-state solution
reached after 10000 round-trips is considered as the initial
condition for the statistics gathering from the 64 independent
samples of propagations with 10000 round-trips for each set of
parameters of Eq. (\ref{master}). The energy scaling can be provided
by scaling of i) mode-size $S_g$, ii) cavity period $T_{cav}$, and iii) average power $P_{av}$ that affects both $ g_0  = \left(\ell+\kappa \right)\left( {1 +
{{2P_{av} T_{cav} } \mathord{\left/
 {\vphantom {{2P_{av} T_{cav} } {E_s^g }}} \right.
 \kern-\nulldelimiterspace} {E_s^g }}} \right)
$ and $E_s^g$.

\section{Results and discussion}\label{results}

In this section, the stability and the statistic properties of a
mode-locked Yb:YAG thin-disk oscillator will be analyzed in both NDR
and PDR. A circulating pulse will be treated as a dissipative
soliton (DS) of Eq. (\ref{master}). This equation can be considered
as the undistributed generalization of the nonlinear complex
Ginzburg-Landau equation, which is the master equation for modeling
of mode-locked solid-state and fiber oscillators
\cite{kaertner,akh}.

\subsection{Stability threshold in NDR and PDR}\label{stab}

\begin{figure}
\begin{center}
    \includegraphics[width=8.25cm]{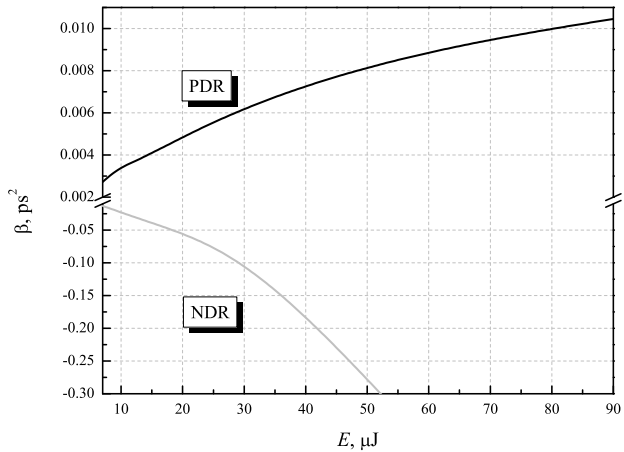}\\
  \caption{Threshold GDD coefficients $\beta$ providing the stable DS for the NDR (gray curve) and the PDR (black curve) in dependence on the output energy $E$. Stable pulses exist above the black curve for the PDR and below the gray curve for the NDR.}\label{fig2}
  \end{center}
\end{figure}

The dependence of the GDD coefficient $\beta$ providing the DS
stability on the output energy $E$ is shown in Fig. \ref{fig2} for
the NDR (gray curve) and the PDR (black curve). The DS is stable
above the black curve for the PDR and below the gray curve for the
NDR.

For the NDR, the numerical threshold GDD (gray curve in Fig.
\ref{fig2}) scales approximately as $\beta$
[ps$^2$]$\approx-$0.0036$(E$[$\mu$J]$)^{1.7}$ , that is slower than
the simple square law suggests (see Sec. \ref{intro}). For the PDR,
the numerical threshold GDD (black curve in Fig. \ref{fig2}) scales
approximately as
$\beta$[ps$^2$]$\approx$0.0011$(E$[$\mu$J]$)^{0.5}$, that is it
obeys the square root law (see Sec. \ref{intro}). The analytical
distributed model of Refs. \cite{kalash1,kalash2,kalash3} predicts
that the threshold intracavity energy $E_{th}^{in}$ in the PDR obeys

\begin{equation}\label{threshold}
    E_{th}^{in}  = \frac{3\gamma }{{\kappa \zeta ^2 c^2 \Omega _g \sqrt \kappa  }} \left[ {\sqrt {\frac{{6\left( {2 - c} \right)}}{c}}  - \frac{{3\left( {2 + c\left( {c - 2} \right)} \right)}}{{\sqrt {1 - c\left( {c - 3} \right)} }}{\mathop{\rm arctanh}\nolimits} \left( {\frac{{\sqrt {6c\left( {2 - c} \right)} }}{{2\sqrt {1 - c\left( {c - 3} \right)} }}} \right)} \right],
\end{equation}
\noindent where $c=\alpha \gamma/\beta\kappa\zeta <$2. However in
the case under consideration (relatively low $\kappa \zeta$ and
large $\gamma$), the last condition is satisfied starting from the
energy $E >$120 $\mu$J and the GDD $\beta>$0.015 ps$^2$.

The destabilization scenario for both NDR and PDR is multiple pulse
generation. For the PDR, the Q-switch mode-locking instability
appears, as well. As one may see from Fig. \ref{fig2}, the GDD
providing the pulse stabilization is substantially lower in the PDR
than that in the NDR and such a difference increases with $E$. That
is the chirped DS is more robust within a whole range of energy than
the chirp-free (Schr\"{o}dinger) soliton. It should be noted, that
even chirped extension of the last one (so-called negative branch of
the chirped DS \cite{kalash1,kalash2,kalash3}) does not provide an
effective energy scaling.

\begin{figure}
\centering \subfigure{\includegraphics[width=8cm]{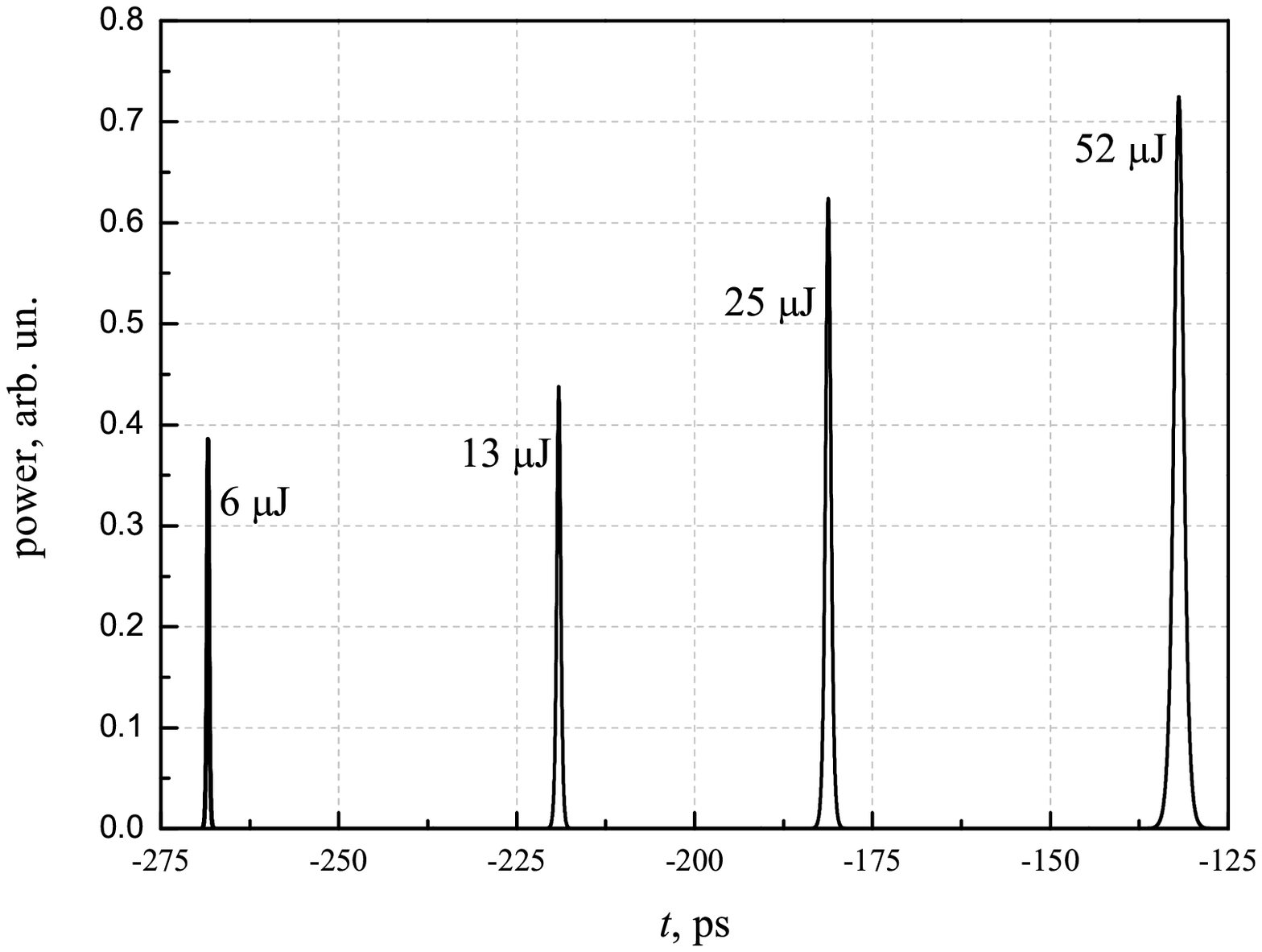}
\includegraphics[width=7.75cm]{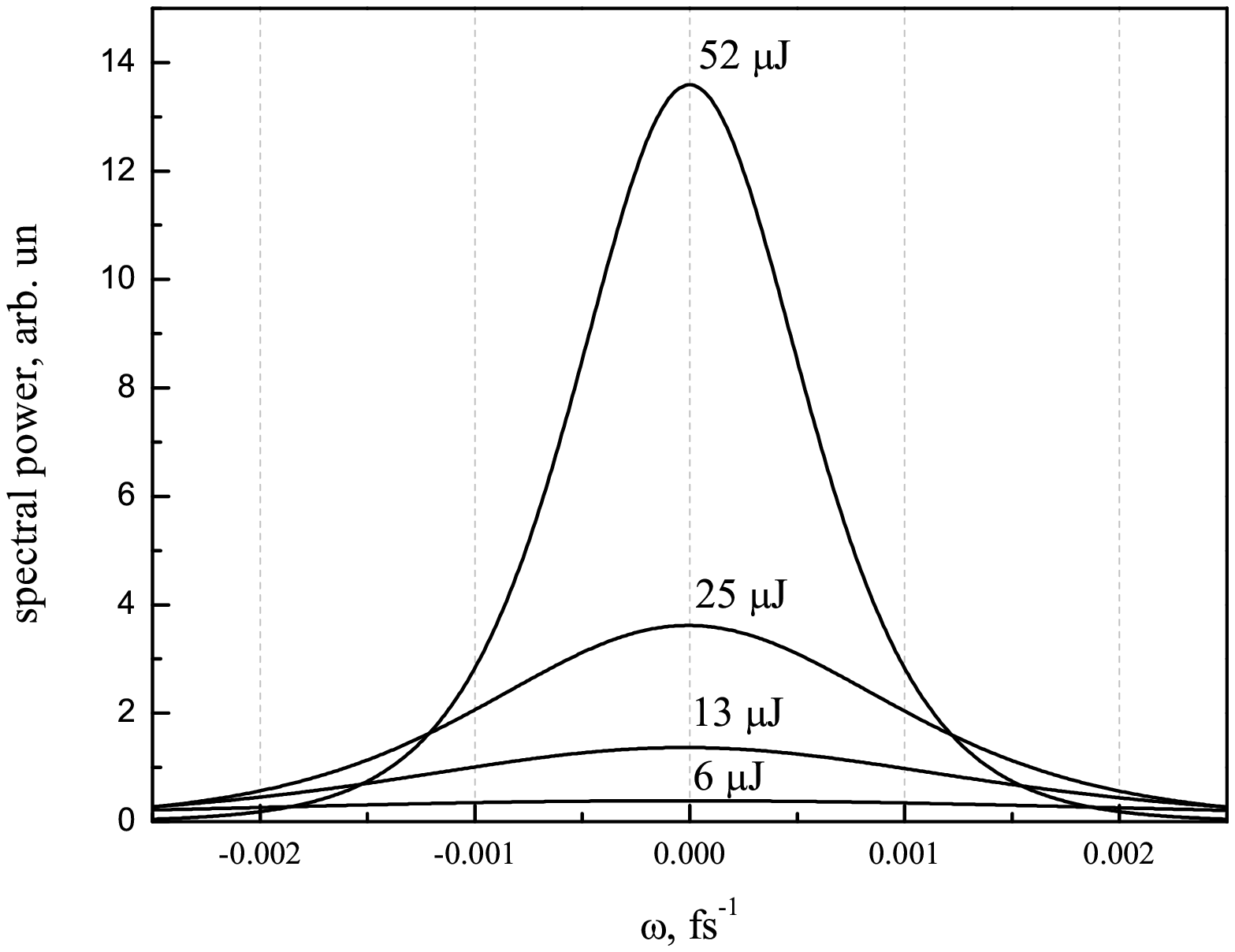}}
  \caption{Instantaneous power $|a(t)|^2$ (left) and spectral power (right) profiles corresponding the CDs at the NDR stability border (Fig. \ref{fig2}). The corresponding energies are superscribed. The frequency $\omega$ corresponds to deviation from the gain maximum.}\label{fig3}
\end{figure}

Fig. \ref{fig3} shows some power and spectral power profiles
corresponding to the DSs at the stability border of the NDR (gray
curve in Fig. \ref{fig2}). One can see, that the pulse duration
increases with energy in agreement with the area theorem: $ T
\propto {{\left| \beta  \right|} \mathord{\left/
 {\vphantom {{\left| \beta  \right|} {E \propto \sqrt {\left| \beta  \right|} }}} \right.
 \kern-\nulldelimiterspace} {E \propto \sqrt {\left| \beta  \right|} }}
$ (for a more careful inspection see next subsection).
Simultaneously, the source of deviation from the square law $E
\propto \beta^2$ is clearly visible: the peak power increases
slightly with energy, as well (Fig. \ref{fig3}, left). In
concordance with the pulse duration growth, the spectral width
decreases with the energy increase (Fig. \ref{fig3}). There exists
no spectral disturbance induced by the dispersion of gain medium
(Eq. (\ref{gain})).

\begin{figure}
\centering \subfigure{\includegraphics[width=8cm]{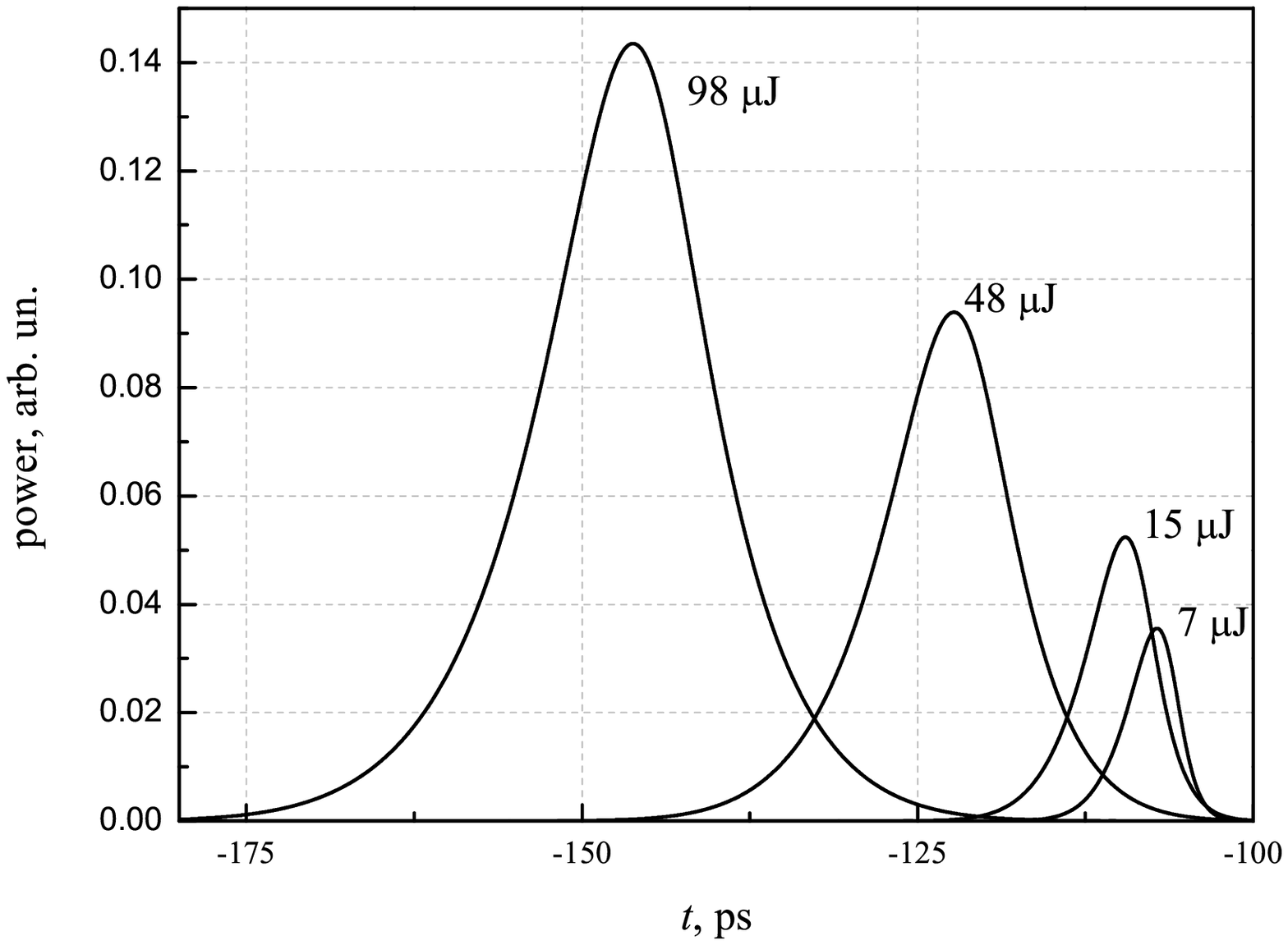}
\includegraphics[width=7.75cm]{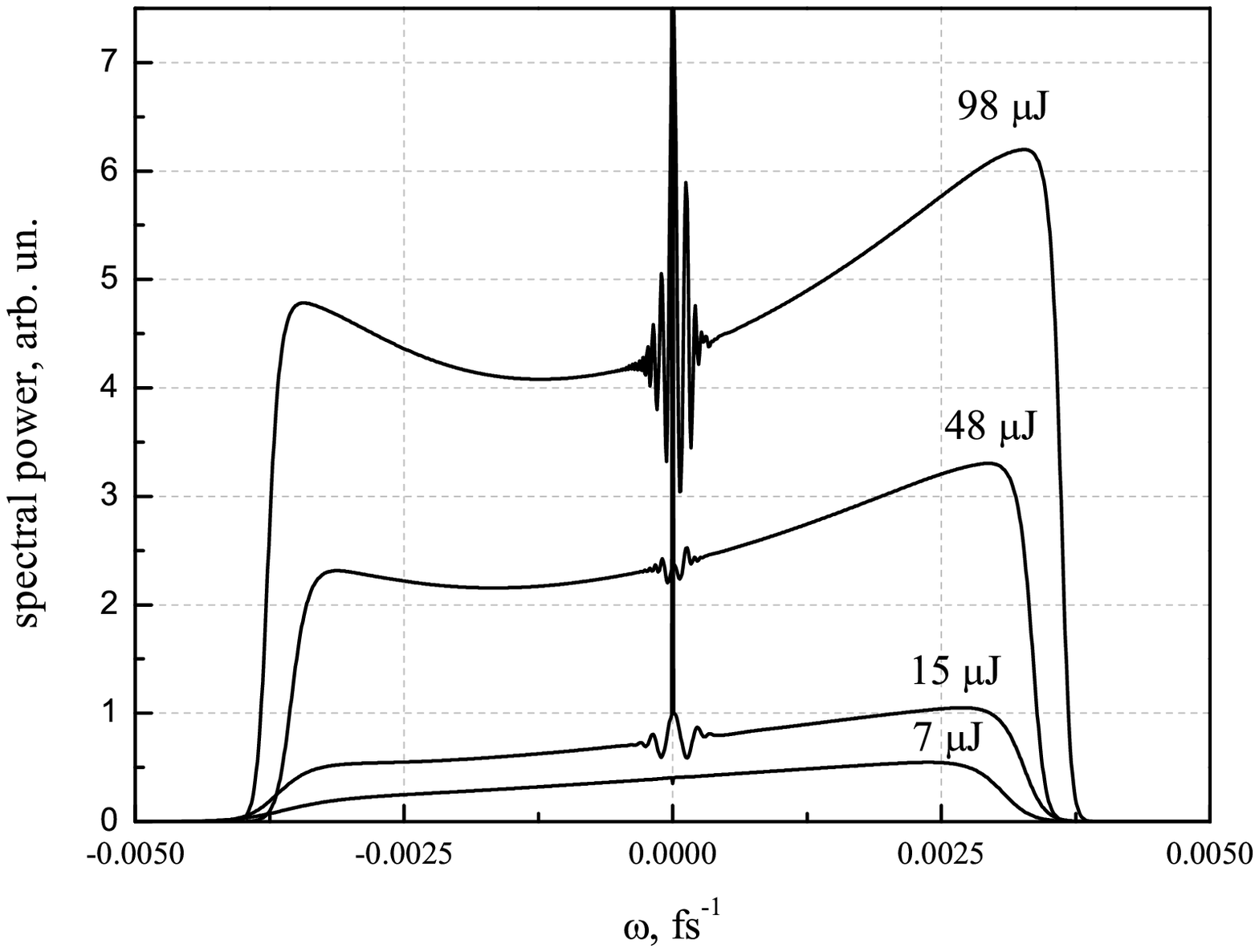}}
  \caption{Instantaneous power $|a(t)|^2$ (left) and spectral power (right) profiles corresponding the chirped CDs at the PDR stability border (Fig. \ref{fig2}). The corresponding energies are superscribed. The frequency $\omega$ corresponds to deviation from the gain maximum.}\label{fig4}
\end{figure}

The analogous profiles for the PDR are shown in Fig. \ref{fig4}. The
DS width, the peak power, and the spectral width $\Delta$ grow with
the energy in the PDR. The analytical model of
Refs.\cite{kalash1,kalash2,kalash3} predicts the following relation
for the last two parameters: $\gamma P_0 = \beta \Delta^2$, and the
$\Delta$-growth with $\beta$ corresponds to the so-called positive
branch of the chirped DS, which exists along the stability
threshold. The analytical theory \cite{kalash1,kalash2,kalash3}
predicts also that $\Delta$ increases with $\beta$ up to some
maximum value of GDD and then the spectrum shortens. But such a
shortening is not possible for the parametrical range under
consideration. It should be noted, that $P_0$ is reduced and
$\Delta$ is enlarged in the PDR in comparison with those parameters
in the NDR.

One may see, that the spectra are asymmetrical in the PDR. Such an
asymmetry has been observed experimentally in the low energy limit
in Ref. \cite{assp}, and we interpret it as a manifestation of the
gain dispersion \cite{kalash5}. Also, some CW-like perturbation is
visible in the vicinity of spectrum center (Fig. \ref{fig4}). This
perturbation appears in the vicinity of stability border
\cite{kalash5}, enhances with $E$, and can be suppressed at higher
GDD. The spectra are truncated, but with smoothed edges. As has been
shown in \cite{kalash1,kalash4}, such a smoothing results from the
relation $1/\Omega_g^2 > \beta$ and $\kappa \zeta \approx \gamma$.
The spectrum becomes concave with sharp edges, when the $E$ (and,
correspondingly, $\beta$) increases.

Figs. \ref{fig3}, \ref{fig4} demonstrate the relative pulse
time-delay, which value changes with $E$ but in opposite directions
for the NDR and the PDR. The group-delay of a DS and its statistical
properties will be considered in the next subsection.

\subsection{Group-delay of a dissipative soliton and its statistical properties}\label{delay}

\begin{figure}
\centering \subfigure{\includegraphics[width=8cm]{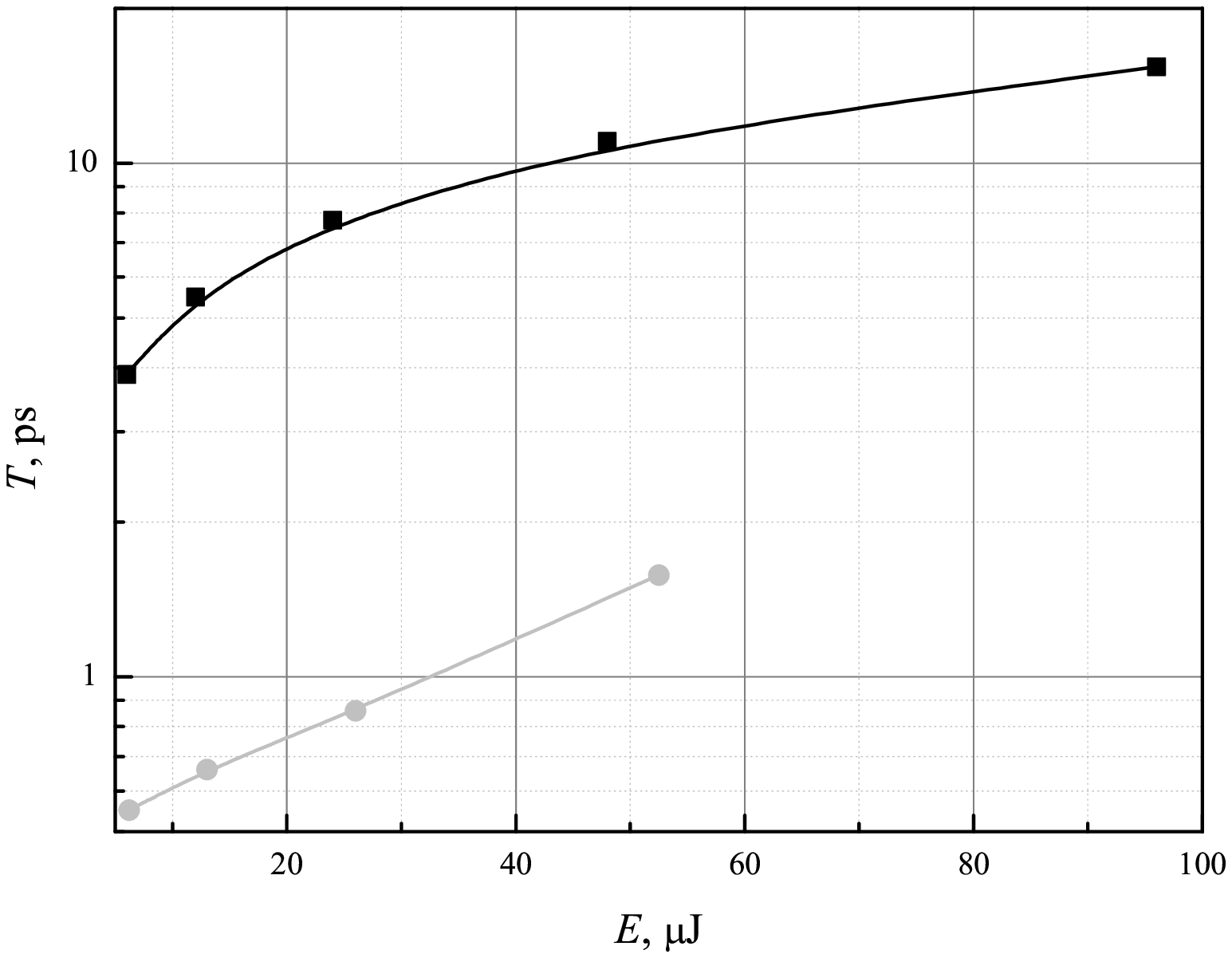}
\includegraphics[width=7.8cm]{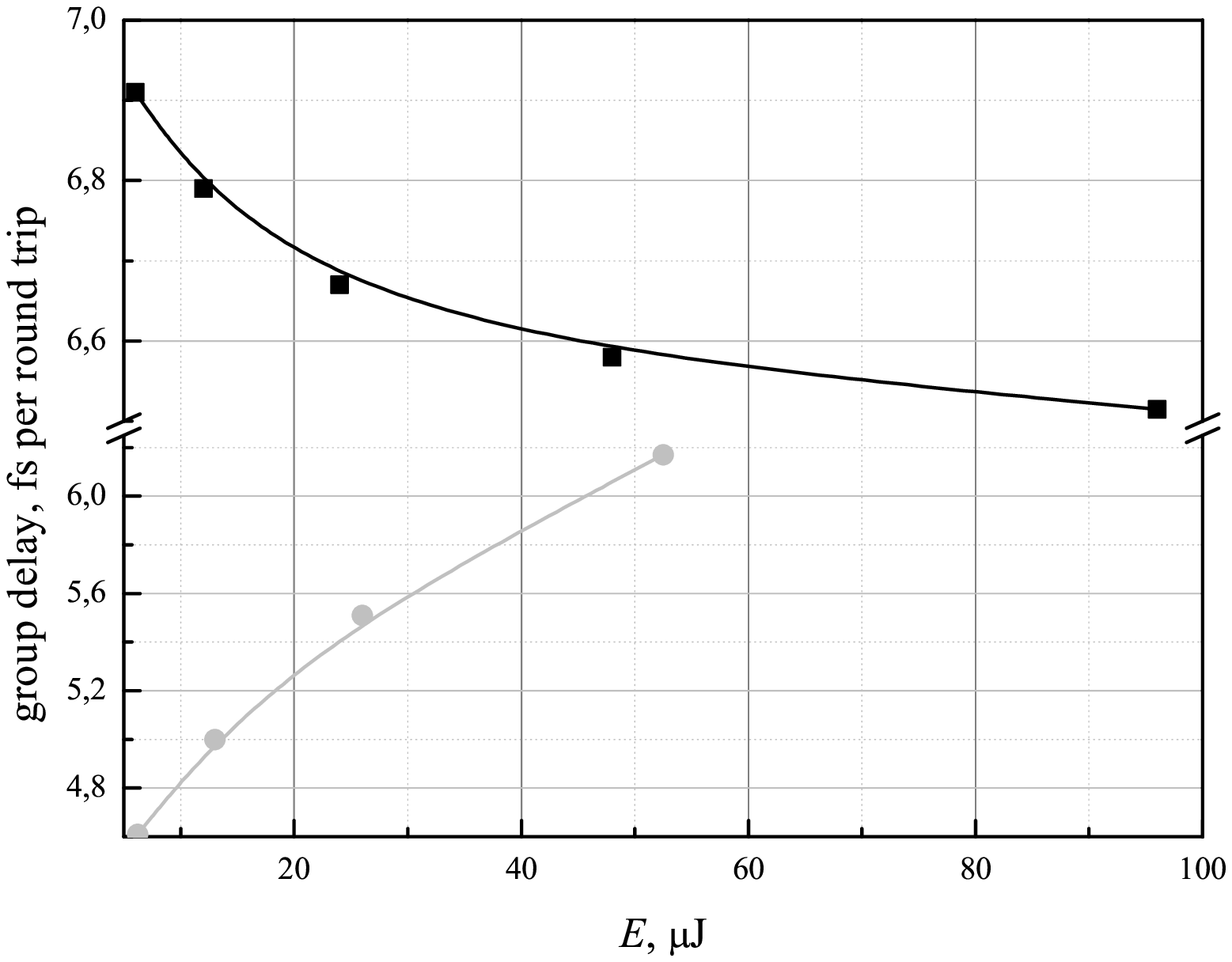}}
  \caption{Pulse duration $T$ (left) and pulse group delay (right) in dependence on the output energy $E$ for the PDR (black squares and lines) and the NDR (gray circles and lines) at the stability border.}\label{fig5}
\end{figure}

The above described increase of $T$ with $E$ is illustrated by Fig.
\ref{fig5}. The durations in the PDR excess those in the NDR and
scale approximately as $T$[ps]$\approx$1.5 $(E$[$\mu$J$])^{0.51}$
(squares and black curve in Fig. \ref{fig5}). In the NDR, the pulse
duration (circles and gray curve in Fig. \ref{fig5}) scales almost
linearly with $E$ (i.e., $T$[ps]$\approx0.36+0.022 E$[$\mu$J]) in
agreement with the area theorem. However, more careful inspection
shows an appearance of higher-order corrections, which enhance the
pulse broadening with $E$ (the best fit is
$T$[ps]$\approx2.4\times10^{-4}(90+ E$[$\mu$J]$)^{2.7}$).

The group delays $\delta$ are shown in Fig. \ref{fig5}, as well. In
the absence of gain dispersion, the group delay equals to zero by
definition of the local time $t$. The gain dispersion induces a
positive delay for both NDR and PDR. However, there is some additive
to this delay in the PDR for shorter pulses (squares and black curve
in Fig. \ref{fig5}). Inversely, the group delay decreases in the NDR
and such a decrease enhances for shorter pulses (circles and gray
curve in Fig. \ref{fig5}). The energy derivatives of $\delta$ are
$d\delta/dE\approx 0.5/E^{0.86}$ and $d\delta/dE\approx -0.16/E$ (in
fs/$\mu$J per round-trip) for the NDR and the PDR, respectively. The
derivative is substantially larger for the NDR therefore one may
expect that the chirped DS is more stable against the timing jitter.
However, such an expectation requires a verification because the
energy $E$ is not independent variable.

The net-variation of $\delta$ can be expressed as

\begin{equation}\label{delay}
   \Delta \delta  \approx \Delta g_0 \left[ {\frac{{\partial \delta }}{{\partial g_0 }} + \frac{{\partial E}}{{\partial g_0 }}\frac{{\partial \delta }}{{\partial E}}} \right],
\end{equation}

\noindent if a source of jitter is assumed to be the gain variations
$\Delta g_0$. In Eq. (\ref{delay}), we divide the total variation of
$\delta$ into two parts: i) variation due to change of an amplitude
of the gain dispersion, and ii) variation due to change of the pulse
characteristics. The last is most interesting because it is a
dynamical, i.e. nonlinear, effect. Since the ``slope efficiency''
${{\partial E} \mathord{\left/
 {\vphantom {{\partial E} {\partial g_0  \approx 10E}}} \right.
 \kern-\nulldelimiterspace} {\partial g_0  \approx 10E}}
$ [$\mu$J] for both regimes under consideration, the
$\delta$-variation has to be reduced in the PDR: $ \left(
{{{\partial E} \mathord{\left/
 {\vphantom {{\partial E} {\partial g_0 }}} \right.
 \kern-\nulldelimiterspace} {\partial g_0 }}} \right)\left( {{{\partial \delta } \mathord{\left/
 {\vphantom {{\partial \delta } {\partial E}}} \right.
 \kern-\nulldelimiterspace} {\partial E}}} \right) \approx
-$1.6 fs per round-trip
 in the PDR versus $
\left( {{{\partial E} \mathord{\left/
 {\vphantom {{\partial E} {\partial g_0 }}} \right.
 \kern-\nulldelimiterspace} {\partial g_0 }}} \right)\left( {{{\partial \delta } \mathord{\left/
 {\vphantom {{\partial \delta } {\partial E}}} \right.
 \kern-\nulldelimiterspace} {\partial E}}} \right) \approx
$5$(E$[$\mu$J]$)^{0.14}$ fs per round-trip  in the NDR.

The experimental results concerning the timing jitter are discrepant
\cite{paschotta3,ilday,nature}. On the one hand,  the relative intensity noise level
is reduced for the PDR. On the other hand, such a reduction in Ref.
\cite{ilday} can result from lower loss at the grating compressor in
the PDR. Therefore, a further analysis of the noise properties of
both PDR and NDR is required.

Our numerical modeling is based on the model of Sec. \ref{model}
with the following noise source included: the gain fluctuation and
the spontaneous emission in an active medium. The first conclusion
obtained from the simulations is that the contribution of quantum
noise is negligible under conditions considered. As the
characteristic quantity, the standard deviation of the pulse
position is chosen: $ \sigma(k)  \equiv \sqrt {{{\sum\limits_{i =
1}^N {\left( {\max \left( {\left| {a_k \left( t \right)} \right|^2 }
\right) - \left\langle {\max \left( {\left| {a_k \left( t \right)}
\right|^2 } \right)} \right\rangle } \right)^2 } } \mathord{\left/
 {\vphantom {{\sum\limits_{i = 1}^N {\left( {\max \left( {\left| {a_k \left( t \right)} \right|^2 } \right) - \left\langle {\max \left( {\left| {a_k \left( t \right)} \right|^2 } \right)} \right\rangle } \right)^2 } } N}} \right.
 \kern-\nulldelimiterspace} N}}
$, where $N=$64 is the number of statistically independent samples
of the steady-state pulse propagations during $k$ cavity
round-trips, $\max \left( {\left| {a_k \left( t \right)} \right|^2 }
\right)$ is the position of the pulse power maximum, and $
\left\langle {...} \right\rangle $ is its mean value. The
simulations demonstrate a linear dependence of $\sigma(k)$ and $
\left\langle \max \left(\left|a_k \left( t \right) \right|^2\right)
\right\rangle $ on $k$.

\begin{figure}
\centering \subfigure{\includegraphics[width=8cm]{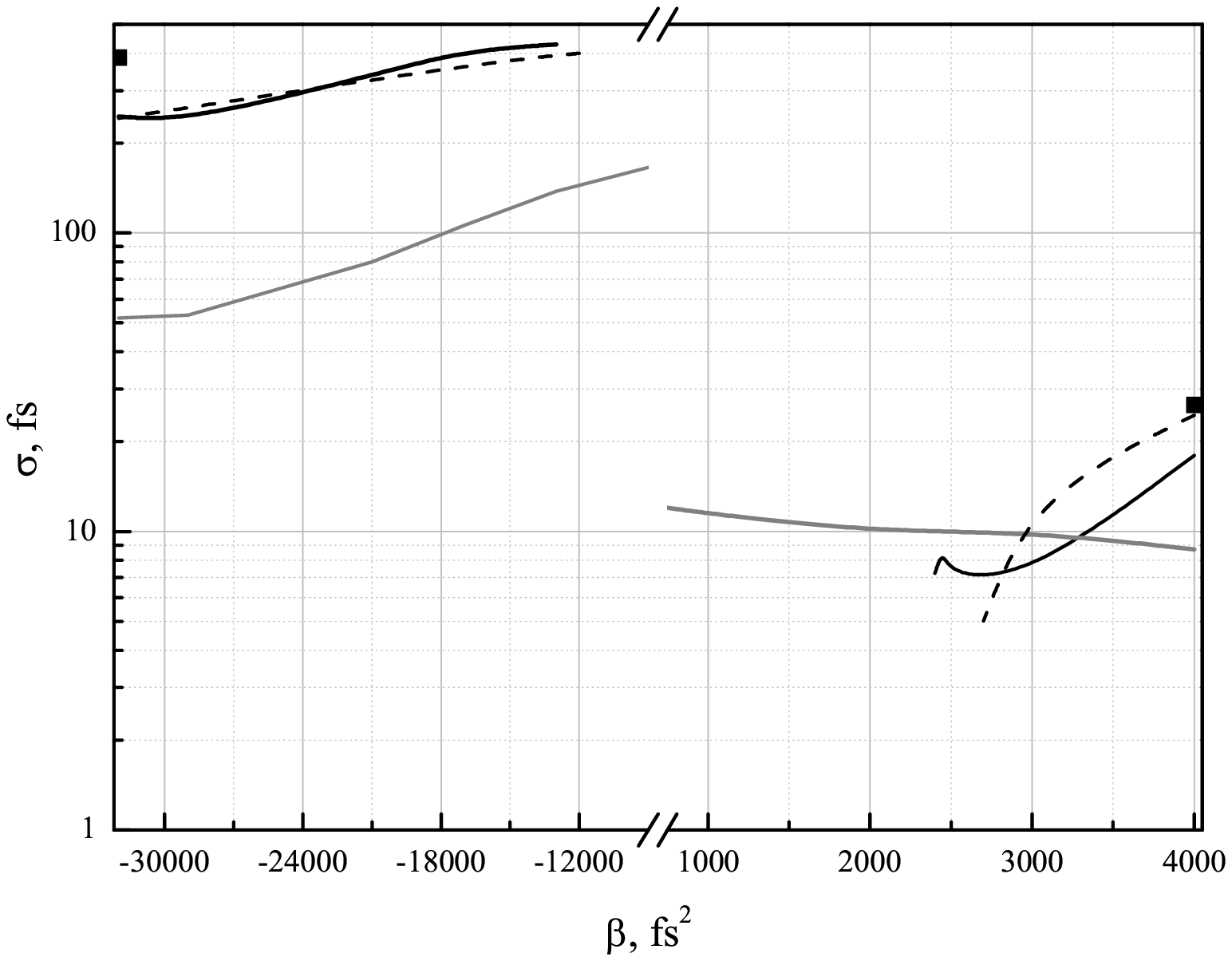}
\includegraphics[width=7.75cm]{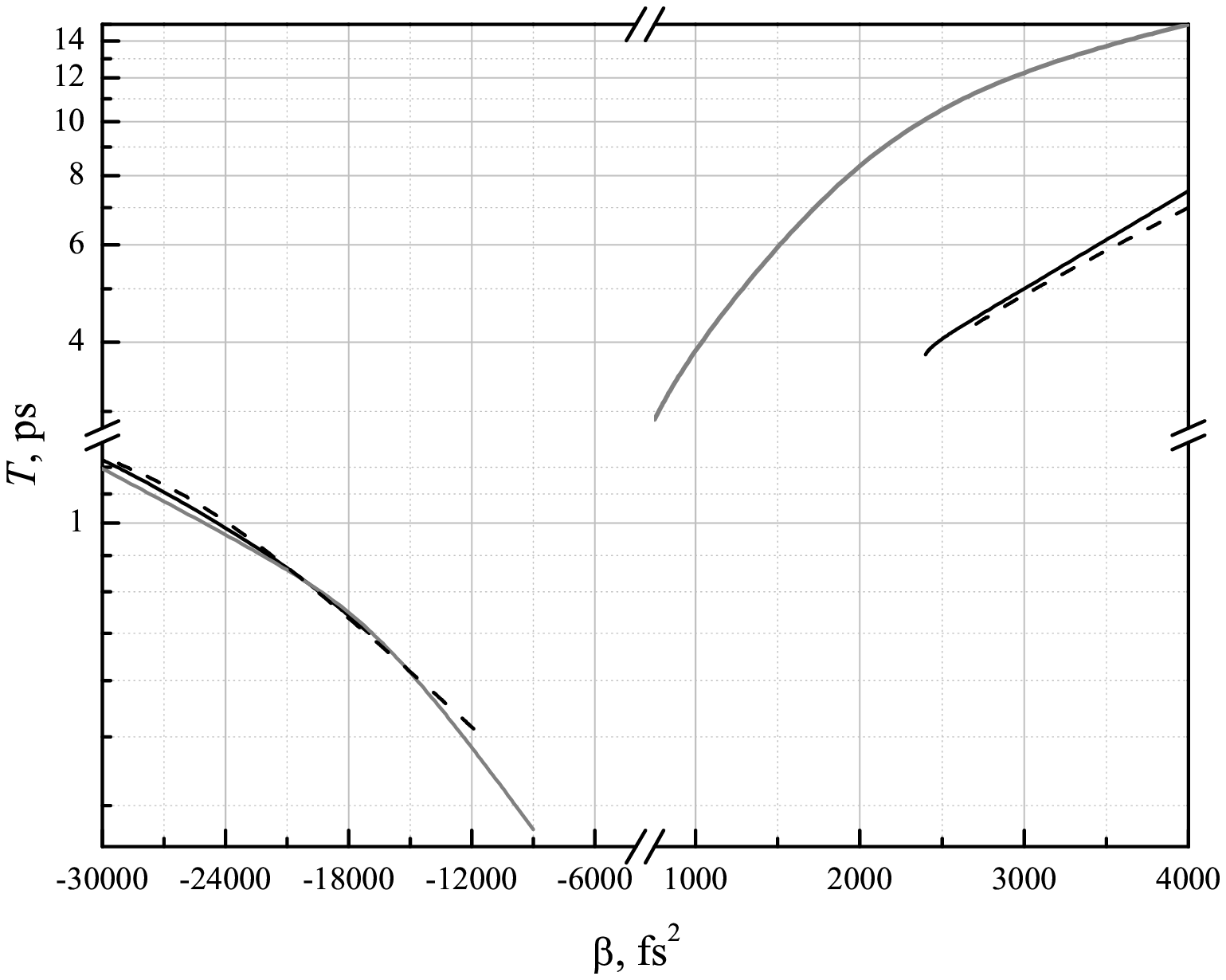}}
  \caption{Jitter parameter $\sigma(k)$ (left) and $T$ (right) corresponding to the stable DS. $k=$10000, $\Omega_g=$5.3 (solid and dashed black curves as well as squares) and 10.5 THz (gray curves). The output energy $E=$6 (dashed curves), 7 (solid curves) and 14 $\mu$J (squares); $\zeta=$0.71 (solid curves and squares) and 0.35 MW$^{-1}$ (dashed curves). The other parameters are given in Sec. \ref{model}.}\label{fig6}
\end{figure}

The dependencies of the group delay standard deviation $\sigma$ and
the DS width $T$ on the dispersion GDD are shown in Fig. \ref{fig6}.
One can see, that the $\sigma$-parameter characterizing the timing
jitter is substantially reduced in the PDR. The dependencies
obtained suggest that the source of such reduction is the negative
passive feedback induced by a spectral dissipation:

1) The chirped DS has a broader spectrum with a concentration of
spectral energy at the spectrum edges, where dissipation is maximum
(Fig. \ref{fig4}). The energy growth broadens the spectrum and
enhances the spectral loss and vice-versa. Thus, the negative
feedback works: energy growth enhances the spectral loss, energy
reduction reduces the spectral loss. As a result, $\sigma$
diminishes (Fig. \ref{fig6}).

2) The GDD growth shortens the spectrum and the spectral loss
decreases. As a result, $\sigma$ increases, that is the negative
feedback disappears (the $\sigma$-value in the PDR tends to that in
the NDR, black curves in Fig. \ref{fig6}).

3) The decrease of spectral dissipation due to gain band broadening
(i.e. the $\Omega_g$-growth; gray curve in Fig. \ref{fig6})
approaches the PDR's $\sigma$ to that in the NDR. The character of
dependence on $\beta$ changes, as well: $\sigma$ decreases with the
$T$-growth. The decrease of $\sigma$ with the $\Omega_g$-growth in
the NDR means that the main source of the group-delay, that is the
gain dispersion, weakens with the gainband broadening. Since the
gain band of a Yb-fiber oscillator is broader than that for a Yb:YAG
solid-state thin-disk oscillator, this conclusion means a reduction
of difference between the noise properties of the PDR and the NDR
for a fiber oscillator \cite{ilday}. Simultaneously, the
$\sigma$-decrease with the pulse duration (or $|\beta|$) growth in
the NDR (Fig. \ref{fig6}) can be explained as a result of
diminishing action of a gain dispersion on the broadened chirp-free
DS (such an action can be approximately described as $
 \propto \left( {1/\Omega _g } \right){d \mathord{\left/
 {\vphantom {d {dt \propto 1/\Omega _g T}}} \right.
 \kern-\nulldelimiterspace} {dt \propto 1/\Omega _g T}}
$ ).

The decrease of the inverse saturation power $\zeta$ does not affect
the stability noticeably  (dashed curves in Fig. \ref{fig6}). The
gainband broadening ($\Omega_g$-growth) enhances the stability
against CW-excitation and multipulsing in the PDR (gray curve in
Fig. \ref{fig6}) because the stability parameter is $\propto \beta
\Omega_g^2$ \cite{kalash1,kalash2}, that is the stabilizing GDD is
inversely proportional to $\Omega_g^2$.

\begin{figure}
\begin{center}
    \includegraphics[width=8cm]{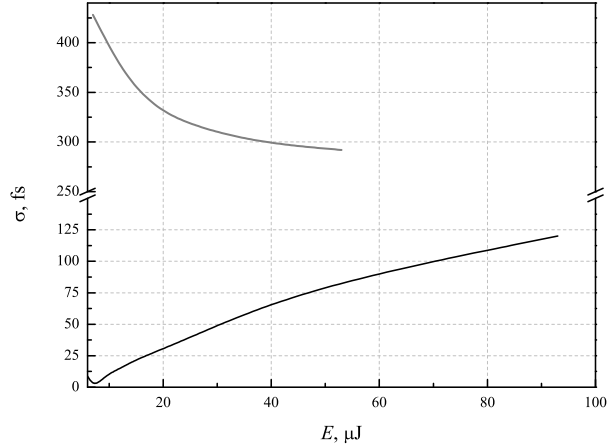}\\
  \caption{Standard deviation $\sigma(k)$ at the stability border (Fig. \ref{fig2}) in dependence on $E$ for the PDR (black curve) and the NDR (gray curve). $k=$10000.}\label{fig7}
  \end{center}
\end{figure}

The energy growth (squares in Fig. \ref{fig6} correspond to the
minimum $|\beta|$ providing the DS stability) increases the
stabilizing GDD (see Fig. \ref{fig2}). The $\sigma$ variation with
$E$ is different for the PDR and the NDR (Fig. \ref{fig7}). The
standard deviation of the pulse maximum location decreases with $E$
for the NDR (the measurement in a 6 $\mu$J Yb:YAG oscillator demonstrates the jitter of 125 fs \cite{ufo}). One may assume that such a decrease results from the conjecture
${{\partial \delta } \mathord{\left/
 {\vphantom {{\partial \delta } {\partial g_0 }}} \right.
 \kern-\nulldelimiterspace} {\partial g_0 }} \propto {1 \mathord{\left/
 {\vphantom {1 T}} \right.
 \kern-\nulldelimiterspace} T}
$ in Eq. (\ref{delay}). It should be repeated, that the noise source considered has not a quantum nature (versus Ref. \cite{paschotta3}, for instance).

The standard deviation $\sigma$ (i.e. the jitter parameter) is
substantially reduced in the PDR (Fig. \ref{fig7}) and equals to few
femtoseconds in the vicinity of $E\approx$7 $\mu$J. We attribute
this reduction to the negative passive feedback caused by the
spectral loss (see above). In the PDR, $\sigma$ increases with $E$
despite the fact that the pulse duration increases, as well (Fig.
\ref{fig5}). Such an effect is not incorporated in Eq. (\ref{delay})
and one can assume, that the source of jitter is connected with a
spectrum broadening and an enhancement of spectral components
located at the spectrum edges (the spectrum becomes more concave for
higher energy, Fig. \ref{fig4}). Hence, the higher-order dispersion
induced by a Lorentz gainband (which contribution in the spectral
domain is $ \propto {1 \mathord{\left/
 {\vphantom {1 {\left( {1 - i{\omega  \mathord{\left/
 {\vphantom {\omega  {\Omega _g }}} \right.
 \kern-\nulldelimiterspace} {\Omega _g }}} \right)}}} \right.
 \kern-\nulldelimiterspace} {\left( {1 - i{\omega  \mathord{\left/
 {\vphantom {\omega  {\Omega _g }}} \right.
 \kern-\nulldelimiterspace} {\Omega _g }}} \right)}}
$, Eq. (\ref{gain})) affects the DS group delay. This effect will be
analyzed elsewhere.

\begin{figure}
\centering \subfigure{\includegraphics[width=8cm]{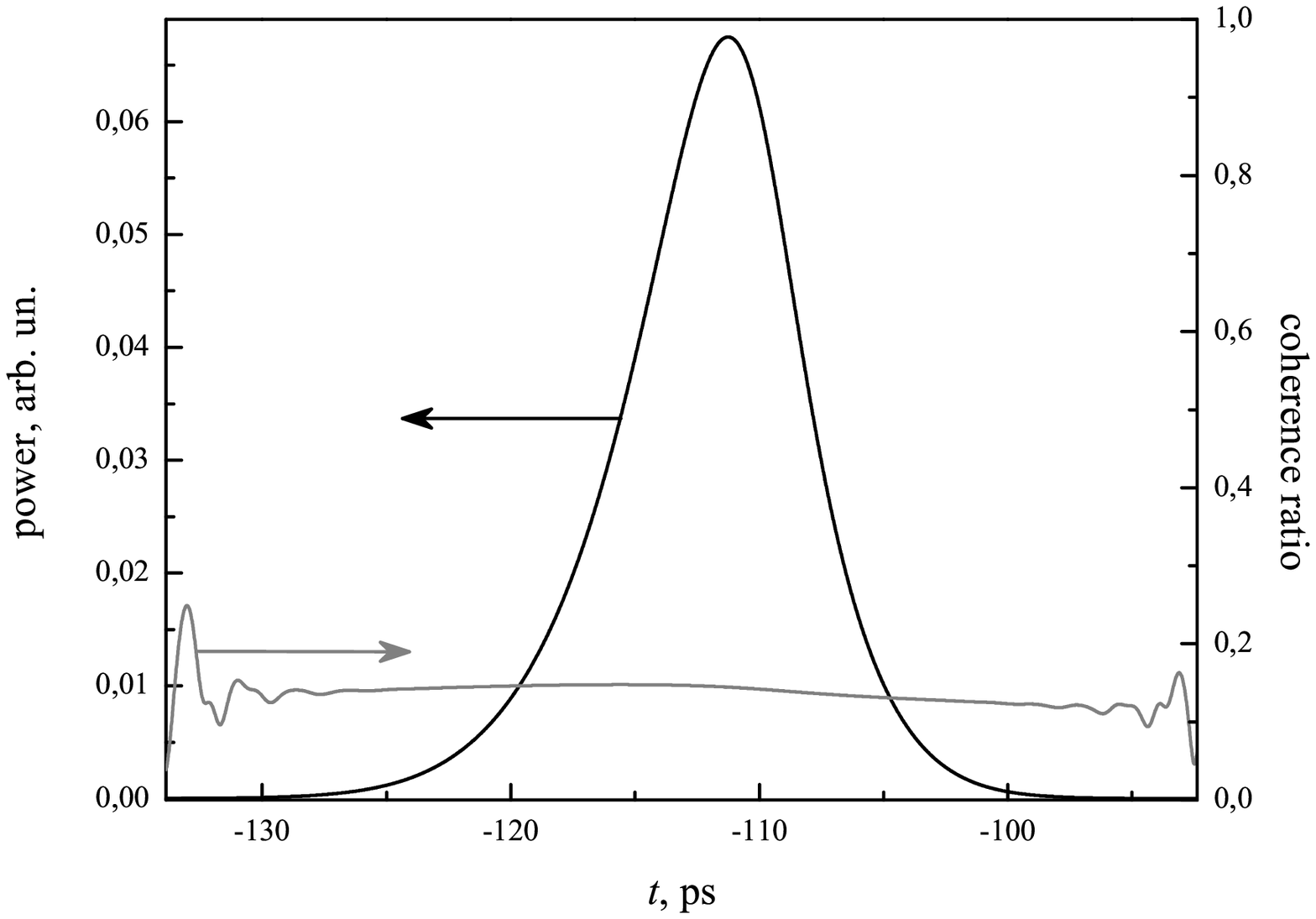}
\includegraphics[width=7.75cm]{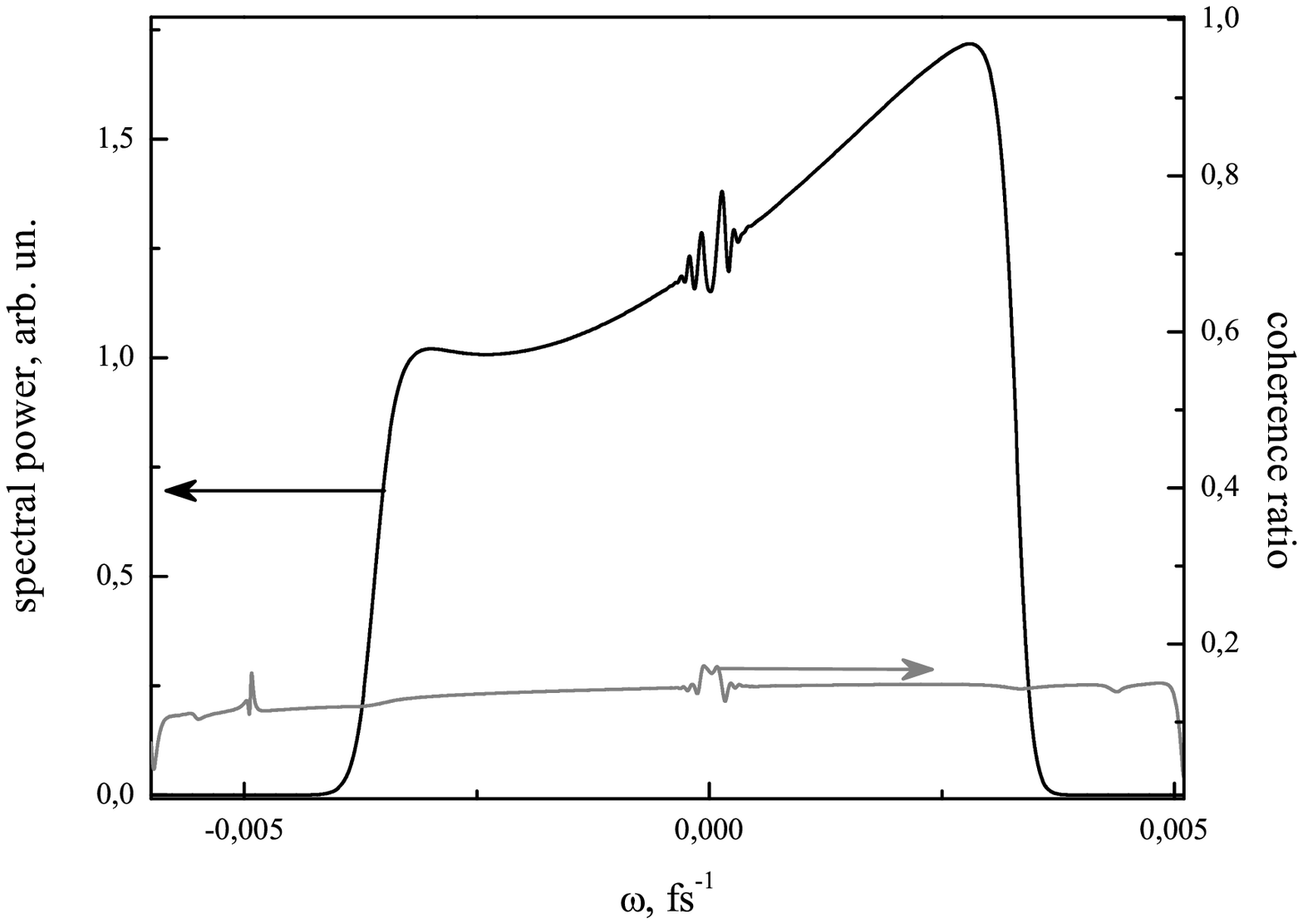}}
  \caption{Averaged power (left) and spectral power (right) profiles (black curves) as well as temporal (left) and spectral (right) coherence ratios (gray curves) for the PDR at the stability border. $E=$46 $\mu$J, $k=$10000.}\label{fig8}
\end{figure}

\begin{figure}
\centering \subfigure{\includegraphics[width=8cm]{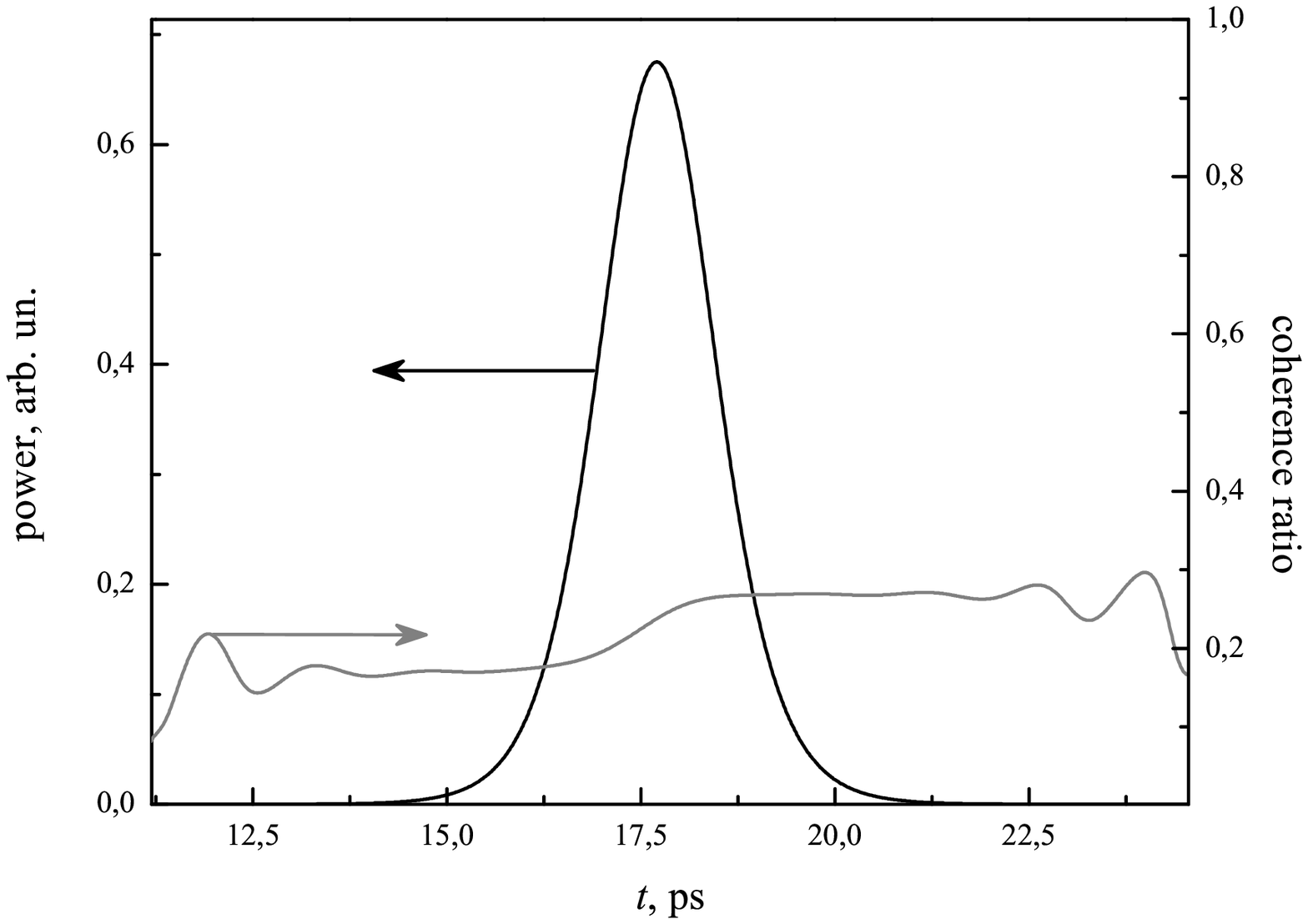}
\includegraphics[width=7.75cm]{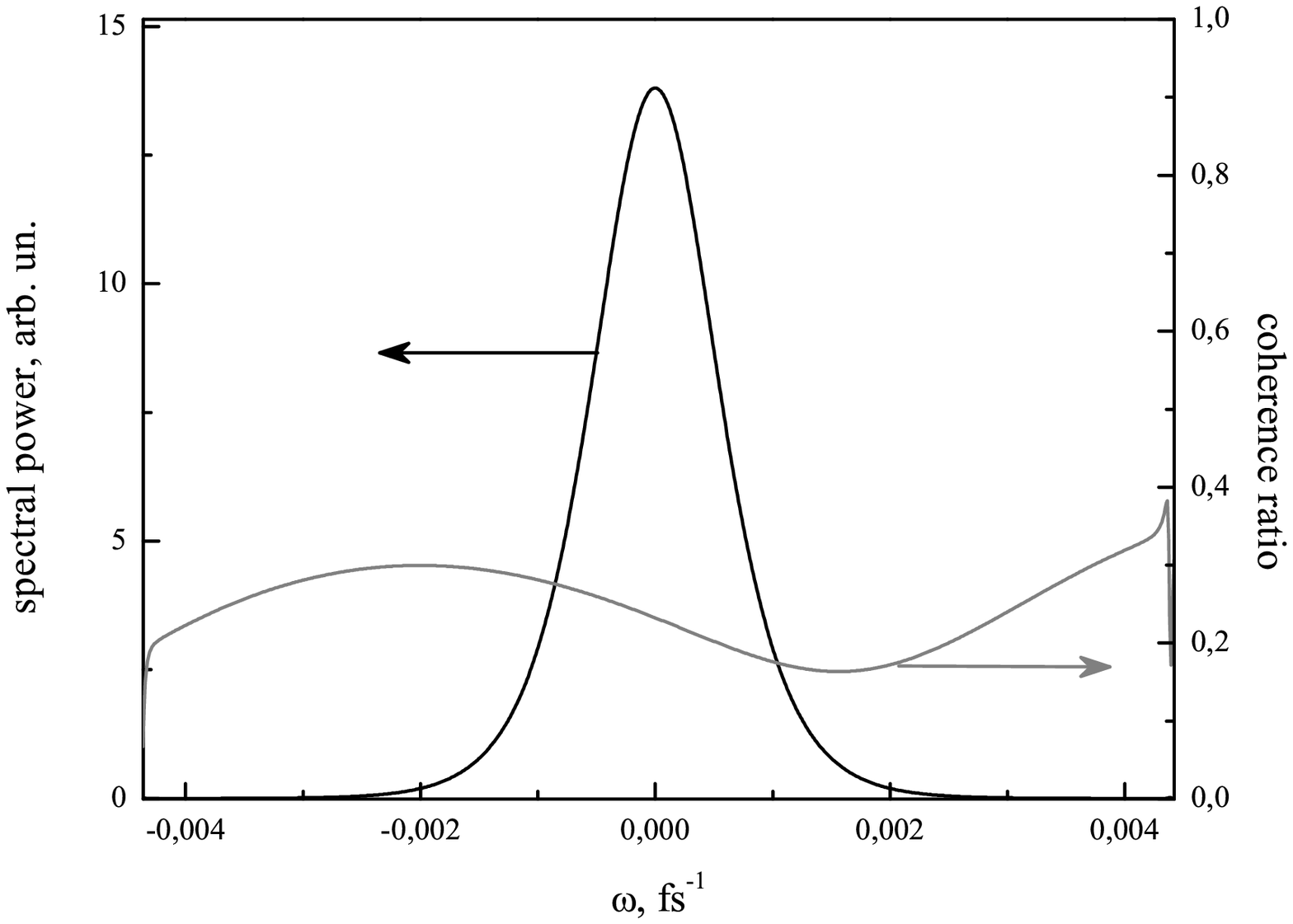}}
  \caption{Averaged power (left) and spectral power (right) profiles (black curves) as well as temporal (left) and spectral (right) coherence ratios (gray curves) for the NDR at the stability border. $E=$53 $\mu$J, $k=$10000.}\label{fig9}
\end{figure}

The timing jitter produced by the gain fluctuations acts
destructively on the DS coherence. Figs. \ref{fig8} and \ref{fig9}
show the averaged power and spectral power profiles as well as the
corresponding coherence ratios for the 64 independent propagation
samples after $k=$10000 cavity round-trips. As the definition of the
coherence ratio, we use \cite{dudley}

\begin{equation}\label{coherence}
    \Gamma  \equiv \frac{{\left| {\left\langle {a_i^* a_j } \right\rangle _{i \ne j} } \right|}}{{\sqrt {\left\langle {\left| {a_i } \right|^2 } \right\rangle _i \left\langle {\left| {a_j } \right|^2 } \right\rangle _j } }},
\end{equation}

\noindent where 64 independent propagation samples after 10000
cavity round-trips are deviled into ($i,j$)-pairs ($i$ and $j$ range
from 1 to 32). $\langle...\rangle$ means averaging over the
corresponding index. $a$ is the complex amplitude in the time or
frequency domains.

One can see, that the coherence ratio (both temporal and spectral)
is substantially below 1 for high-energy pulses. The NDR provides a
higher coherence at the average than that in the PDR. A possible
explanation is that the DS in the PDR is chirped and the power
fluctuations distort the pulse phase that reduces the coherence.
Simultaneously, the coherence ratio is almost constant inside a
chirped DS in both time and frequency domains.

\subsection{Analytical estimations}\label{analytics}

Complicate dynamics of the high-energy oscillators needs the cumbersome numerical simulations. Especially, gain saturation, gain dispersion, SPM and higher-order dispersions entangle the contributions from different noise sources and add the correlations between them. As will be shown below, the simple improvement of theory developed in Ref. \cite{paschotta2} allows comparison between the quantum noise limit and the timing jitter caused by gain dispersion for both PDR and NDR.

The intracavity power density of the timing noise induced by a spontaneous emission and a dispersion-mediated effect of frequency fluctuations is \cite{haus2,paschotta2}

\begin{equation}\label{noise}
    S\left( f \right) = \left( {\frac{{2\beta }}{{fT_{cav} }}} \right)^2 \frac{{4\left( {\ell  + \kappa } \right)}}{{\left( {2\pi f} \right)^2  + \tau _g^{ - 2} }}\frac{{h\nu }}{{E^2 T_{cav} }}\int {\omega ^2 \left| {\tilde a\left( \omega  \right)} \right|^2 d\omega }  + \frac{{4\left( {\ell  + \kappa } \right)}}{{\left( {2\pi f} \right)^2 }}\frac{{h\nu }}{{E^2 T_{cav} }}\int {t^2 \left| {a\left( t \right)} \right|^2 dt},
\end{equation}
\noindent where $f$ is the noise frequency and $\tilde a\left( \omega  \right)$ is the Fourier image of the field. It is assumed, that the saturated gain coefficient equals to the net-loss coefficient, $E$ is the intracavity energy. The impact of a gainband as a spectral filter is defined by the parameter $\tau _g  = \frac{{T_{cav} {{\Omega _g^2 } \mathord{\left/
 {\vphantom {{\Omega _g^2 } {\Delta ^2 }}} \right.
 \kern-\nulldelimiterspace} {\Delta ^2 }}}}{{16\left( {\ell  + \kappa } \right)}}
$ ($\Delta$ is the pulse spectral width for the truncated spectrum in the PDR or $1.763/\sqrt{3} T$ for the NDR). Integral $
{{\int {\omega ^2 \left| {\tilde a\left( \omega  \right)} \right|^2 } d\omega } \mathord{\left/
 {\vphantom {{\int {\omega ^2 \left| {\tilde a\left( \omega  \right)} \right|^2 } d\omega } {E \approx \Delta ^2 }}} \right.
 \kern-\nulldelimiterspace} {E \approx \Delta ^2 }}
$ for the PDR or $\approx 0.265 \left(0.315/T \right)^2$ for the NDR. Integral $
{{2\int {t^2 \left| {a\left( t \right)} \right|^2 } dt} \mathord{\left/
 {\vphantom {{2\int {t^2 \left| {a\left( t \right)} \right|^2 } dt} {E \approx 0.529T^2 }}} \right.
 \kern-\nulldelimiterspace} {E \approx 0.529T^2 }}$ for both regimes.

\begin{figure}
\centering \subfigure{\includegraphics[width=8cm]{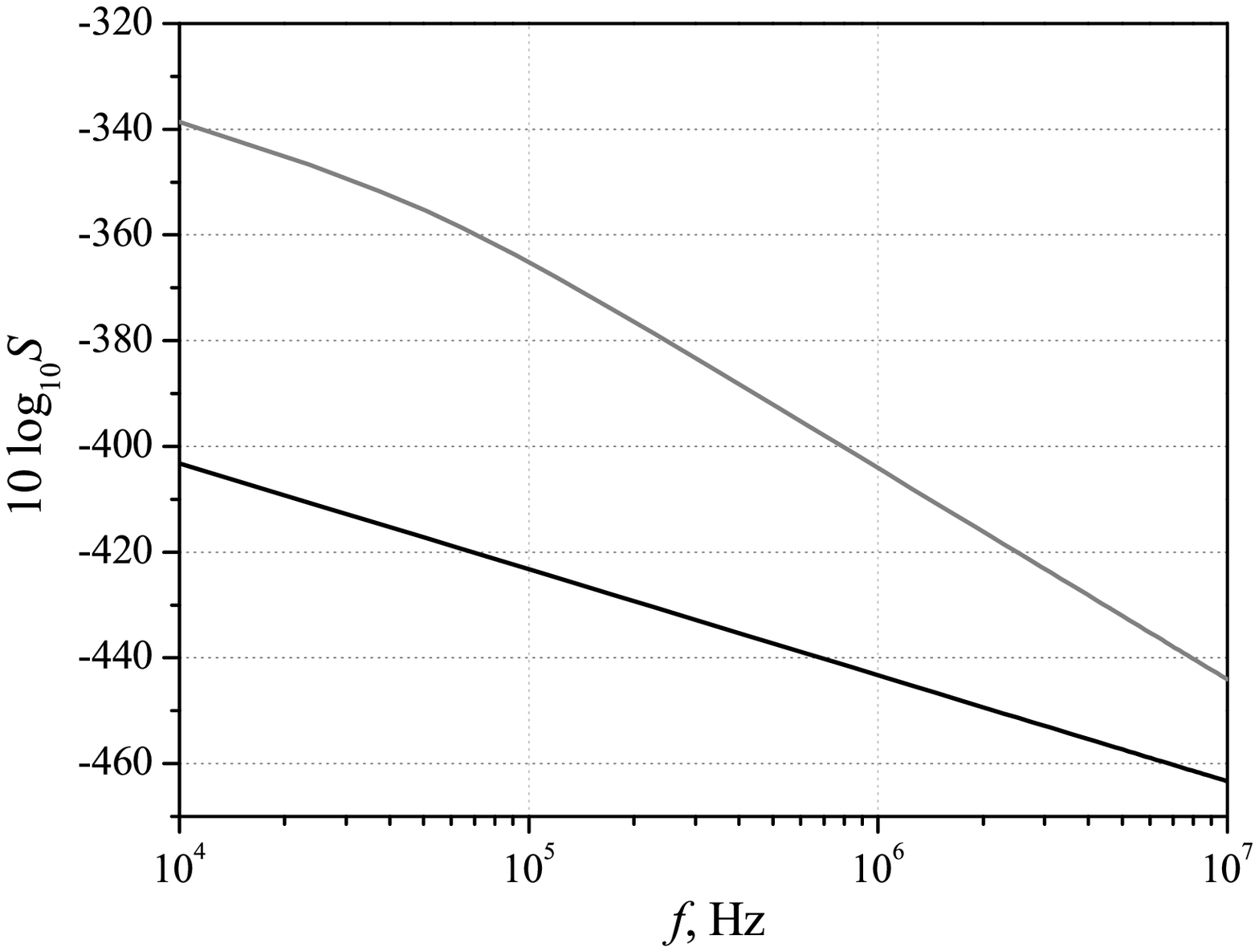}
\includegraphics[width=7.75cm]{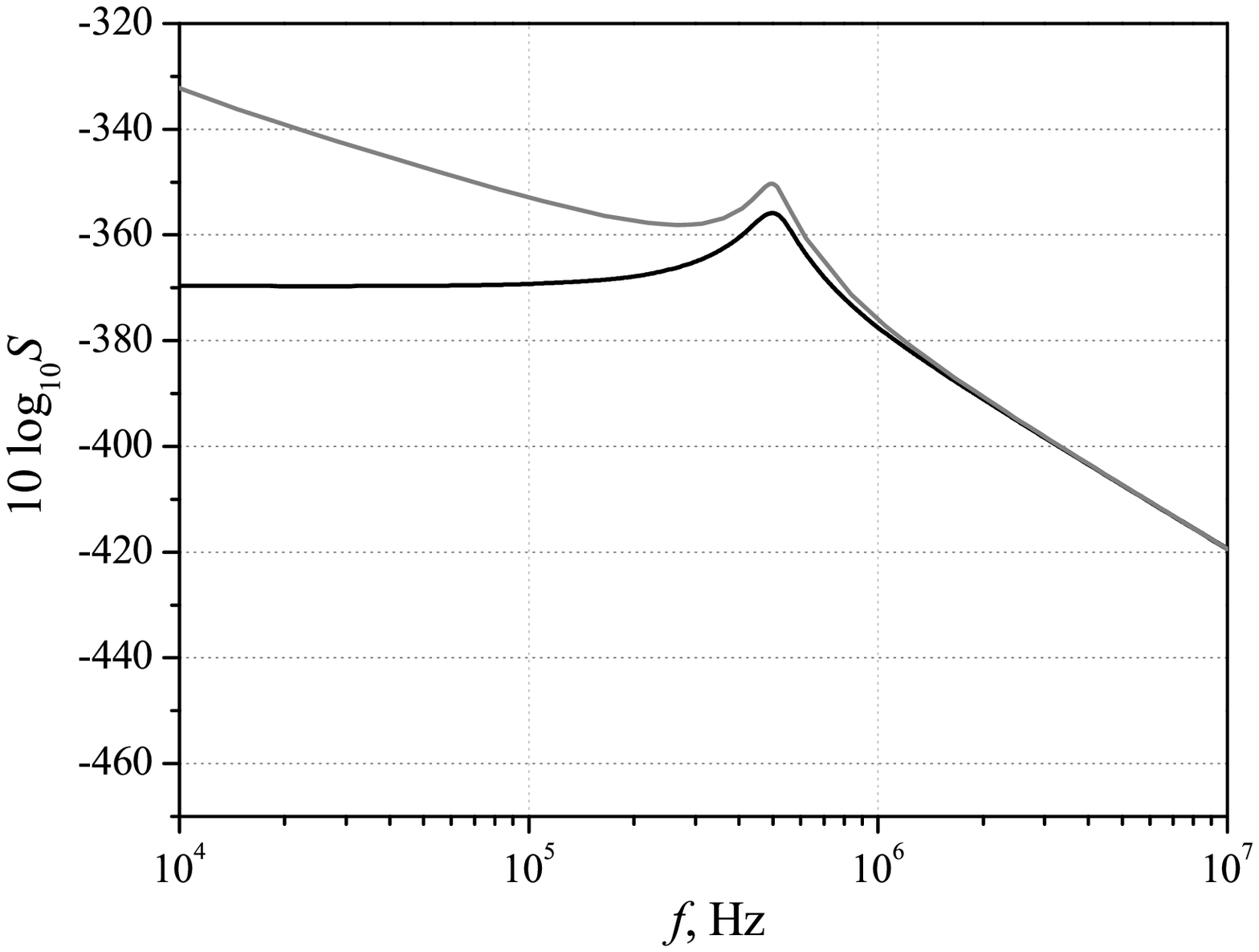}}
  \caption{Quantum limit (left) for the timing noise. The NDR (gray curve) corresponds to $T=$1 ps, $\beta=$-0.01 ps$^2$, the PDR (black curve) corresponds to $\Delta=3$ ps$^{-1}$, $T=$3 ps, $\beta=$0.0025 ps$^2$. $E=$50 $\mu$J, $\Omega_g=$5.3 THz. Timing noise induced by gain fluctuation (right) in the PDR (black curve) and the NDR (gray curve), $\Omega=2\pi \times$50 kHz, $\Theta=2\pi \times$5 kHz.}\label{fig10}
\end{figure}

Left Fig. \ref{fig10} shows the noise spectrum in the quantum limit of (\ref{noise}) for the PDR (black curve) and the NDR (gray curve). One can see, that the NDR has substantially higher noise, especially in the low-frequency range. Analysis of Eq. (\ref{noise}) demonstrates that such an excess results due to the strong noise suppression of frequency fluctuations in the PDR (first term in Eq. (\ref{noise})). This results from i) more effective spectral filtering for the broad truncated spectrum of the chirped DS and ii) lower $|\beta|$ in the PDR. On the other hand, the timing effect of spontaneous emission is stronger in the PDR due to larger $T$ (second term in Eq. (\ref{noise})). As a result, the difference between noise levels decreases with $f$.

As it has been shown in the previous subsection, the gain dispersion translates the gain fluctuations into the timing noise. The trivial correction of the model presented in Ref. \cite{paschotta2} with taking into account the negative feed-back induced by spectral filtering results in

\begin{equation}\label{gn}
  S\left( f \right) = \frac{1}{{\left( {2\pi f} \right)^2  + \tau _g^{ - 2} }}\left( {\frac{1}{{T_{cav} \Omega _g }}} \right)^2 S_g \left( f \right),
\end{equation}
\noindent where $S_g$ is the gain noise. Let's $S_g$ results from quasi-harmonic gain oscillations with the amplitude $0.05\left(\ell + \kappa\right)$, the frequency $\Omega$ and the Lorentz bandwidth $\Theta$.  Then

\begin{equation}\label{g}
    S_g \left( f \right) = \left[ {0.05\left(\ell +\kappa \right)} \right]^2 \sqrt {\frac{2}{\pi }} \frac{{\Theta \left[ {\Theta ^2  + \left( {2\pi f} \right)^2  + \Omega ^2 } \right]}}{{\left( {2\pi f} \right)^4  + \left( {\Theta ^2  + \Omega ^2 } \right)^2  + 2\left( {2\pi f} \right)^2 \left( {\Theta ^2  - \Omega ^2 } \right)}}
\end{equation}
\noindent and the net-noise spectra in the NDR (gray curve) and the PDR (black curve) are shown in Fig. \ref{fig10} (right). From comparison of the left and right Figs. 10, one can see that the noise exceeds the quantum limit in the vicinity of the resonant peak, which corresponds to the frequency of the gain modulation. The low-frequency branch demonstrates the noise-excess in the NDR (gray curve, right Fig. \ref{fig10}) in comparison with the PDR (black curve, right Fig. \ref{fig10}). Simultaneously, the noise in the PDR with oscillating gain exceeds the quantum limit substantially.

\begin{figure}
\centering \subfigure{\includegraphics[width=8cm]{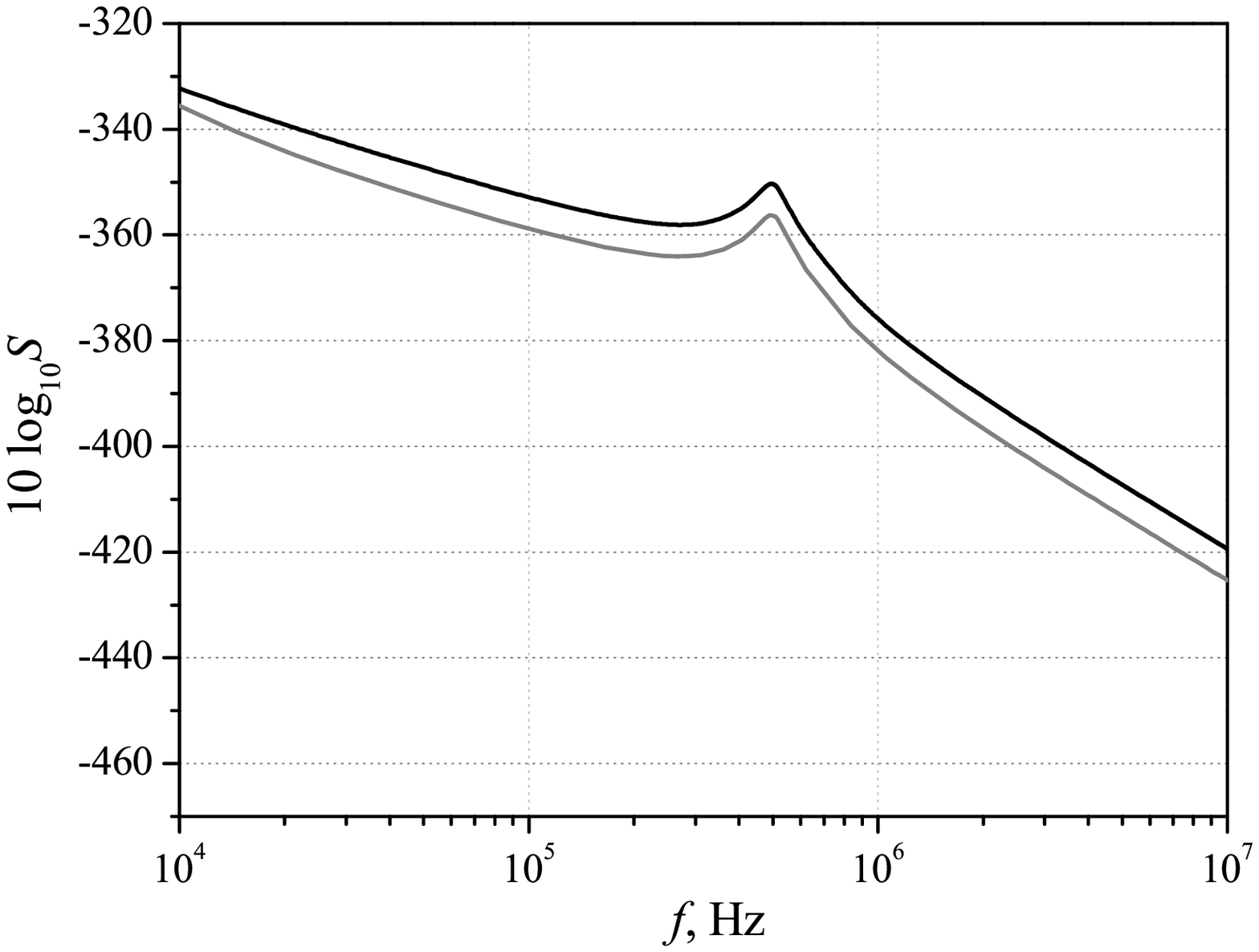}
\includegraphics[width=7.75cm]{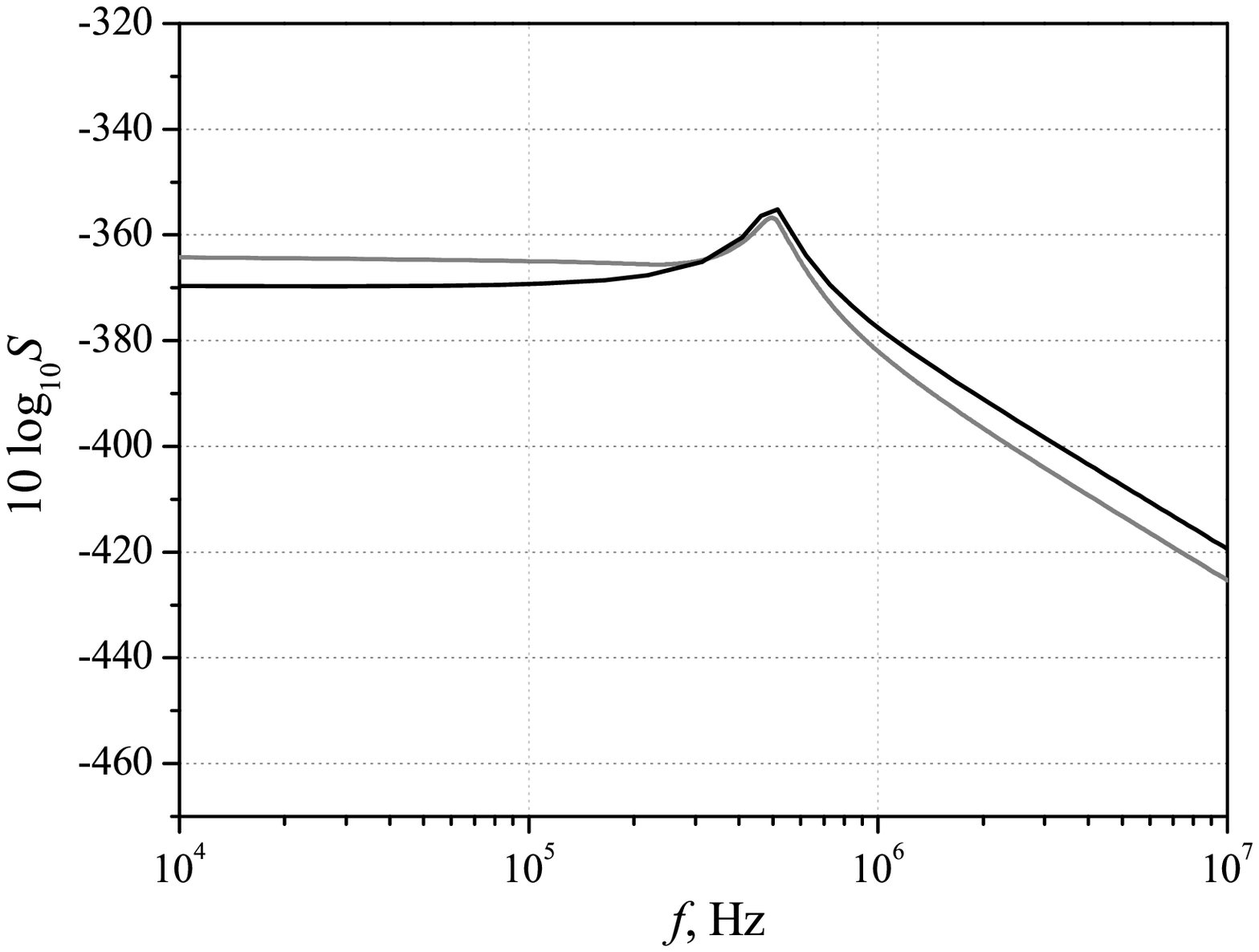}}
  \caption{Noise spectrum induced by gain fluctuation in the NDR (left) and the PDR (right). $\Omega_g=$5.3 THz (black curves) and 10.5 THz (gray curves). Other parameters correspond to Fig. \ref{fig10}.}\label{fig11}
\end{figure}

The gainband broadening, which damps the gain dispersion, reduces the noise in the NDR (left Fig. \ref{fig11}, compare the gray and black curves as well as see left Fig. \ref{fig6}). In the PDR, the gainband broadening enhances the low-frequency noise but reduces the high-frequency noise (the right Fig. \ref{fig11}, compare the gray and black curves as well as see the left Fig. \ref{fig6}).

\section{Conclusions}

The numerical analysis of a mode-locked Yb:YAG thin-disk oscillator
operating in both PDR and NDR is presented. The energy scalability
within a broad energy range (from 6 to 100 $\mu$J) is analyzed. It
is found, that the level of stabilizing GDD is substantially reduced
in the PDR ($\approx$0.003--0.01 ps$^2$ vs. $-$0.02--$-$0.3 ps$^2$
in the NDR). The pulse duration in the PDR is approximately tenfold
of that in the NDR, but the spectra in the PDR is broader. These
spectra is truncated, asymmetrical and concave. In spite of the
pulse duration growth, the spectra broaden with energy for a chirped
pulse.

The timing jitter of a mode-locked Yb:YAG thin-disk oscillator is
analyzed numerically. As a source of jitter, the gain fluctuations
are considered. It is found, that the timing jitter is substantially
reduced in the PDR. It is assumed, that the mechanism of such
reduction is a negative passive feedback induced by spectral
dissipation. We see,  that, in general, the difference between the
levels of timing jitter in the PDR and the NDR decreases with a
gainband broadening, i.e. with transit from a solid-state to a fiber
oscillator. Scaling properties of timing jitter depend on the
regime: the jitter increases with energy for the PDR and decreases
for the NDR. The dependence on dispersion differs, as well: in spite
of the NDR, the timing jitter grows with dispersion in the PDR. The
temporal and spectral coherence is reduced due to jitter for both
regimes, but the reduction is lower for a chirped pulse.

Our simple analytical analysis demonstrates that the quantum timing noise is substantially reduced in the PDR due to suppression of the frequency fluctuations. The difference between the noise levels of the PDR and the NDR decreases with frequency because the timing noise due to spontaneous emission is higher for the PDR as a result of larger pulse duration. The analysis confirms the numerical result that the timing jitter induced by gain fluctuations is suppressed in the PDR in comparison with the NDR for low-frequencies of the noise. The gainband broadening reduces this noise in the NDR and in the PDR (for high-frequencies). Simultaneously, the low-frequency noise induced by gain fluctuations in the PDR increases with the gain broadening.

\acknowledgments The work was supported by Austrian Fonds zur
F\"{o}rderung der wissenschaftlichen Forschung (project P20293) and
Munich Centre for Advanced Photonics (MAP).

\end{document}